\documentclass[12pt]{article}
 \usepackage{graphicx}
 \usepackage[cp1251]{inputenc}

\begin{document}
 \noindent {\footnotesize\it
   Astronomy Letters, 2021, Vol. 47, No 7, pp. 454--473.}
 \newcommand{\dif}{\textrm{d}}

 \noindent
 \begin{tabular}{llllllllllllllllllllllllllllllllllllllllllllll}
 & & & & & & & & & & & & & & & & & & & & & & & & & & & & & & & & & & & & & &\\\hline\hline
 \end{tabular}

  \vskip 0.5cm
\centerline{\bf\large Study of the Influence of an Evolving Galactic Potential}
\centerline{\bf\large on the Orbital Properties of 152 Globular clusters}
\centerline{\bf\large with Data from the Gaia EDR3 Catalogue}
   \bigskip
   \bigskip
  \centerline
 {A. T. Bajkova \footnote [1]{e-mail: bajkova@gaoran.ru}, A. A. Smirnov and V. V. Bobylev}
   \bigskip

  \centerline{\small\it Pulkovo Astronomical Observatory, Russian Academy of Sciences,}

  \centerline{\small\it Pulkovskoe sh. 65, St. Petersburg, 196140 Russia}
 \bigskip
 \bigskip
 \bigskip

{\bf Abstract}---We have studied the influence of an evolving gravitational potential of the Milky Way Galaxy on the orbital motion of 152 globular clusters with proper motions from the Gaia EDR3 catalogue and mean distances from Baumgardt and Vasiliev (2021). To construct a semicosmological evolving model potential with changing masses and sizes of the Galactic components, we have used the algorithm described in Haghi et al. (2015). The adopted axisymmetric three-component model potential of the Galaxy includes a spherical bulge, a flat Miyamoto--Nagai disk, and a spherical Navarro--Frenk--White dark matter halo.
The orbits are integrated backward in time. We compare the orbital parameters of globular clusters derived in static and evolving potentials when integrating the orbits for 5 and 12 Gyr backward. For the first time we have studied the influence of separately a change in the masses and a change in the sizes of the Galactic components. The changes in the masses and sizes of the components are shown to act on the orbital parameters in the opposite way. At small Galactocentric distances this influence is maximally compensated for. The orbits of distant globular clusters and those with a large apocenter distance undergo the biggest
changes. We show that on time scales up to $-5$~Gyr the orbits of globular clusters in the case of a potential with both changing masses and changing sizes of the components undergo, on average, minor changes compared to the case of a static potential. These changes fit into the limits of the statistical uncertainties caused by the errors in the data. So, on these time scales the Galactic potential may be deemed static. We provide tables with the orbital parameters of globular clusters derived in both static and evolving potentials.


 \subsection*{INTRODUCTION}
At present, our Milky Way Galaxy numbers more than 150 globular clusters (GCs) orbiting around the Galactic center and distributed at Galactocentric distances up to 200 kpc. GCs are the oldest objects whose age reaches 13 Gyr and is comparable to the
age of the Universe. Basically, they are witnesses of the earliest galaxy formation epoch and, hence, their investigation is a powerful means of studying the physical conditions in the Universe over the course of its evolution.

A very important aspect of investigating GCs is a study of their orbital history, which, in turn, is determined by the evolution of the Galactic potential.
This evolution depends on many processes, including
secular ones, such as the formation of a bar (Perez-Villegas et al. 2020), and processes attributable to the interaction with other galaxies. These primarily
include satellite galaxies (Garrow et al. 2020).
The interaction with the most massive satellite, the
Large Magellanic Cloud, should be noted specially
(see Battaglia et al. 2021). On the whole, the kinematics
of GCs is also determined by the hierarchical
galaxy formation history (Bekki et al. 2005; Arnold
et al. 2011; Trujillo-Gomez et al. 2021). An important
dynamical feature of GCs is also the relationship
between the cluster tidal radius and total Galactic
potential (King 1962). Such a relationship allows
the mass of the galaxy in which a test GC moves,
including the mass of the Milky Way Galaxy, to be
estimated (Bellazzini 2004; Haghi et al. 2015).

To study the orbital motion of GCs, many authors commonly use a static potential, i.e., they assume that it remains constant when integrating the orbits
(see, e.g., Helmi et al. 2018; Massari et al. 2019;
Bajkova et al. 2020). However, observations show
that the size and mass content of galaxies change
significantly with redshift: the sizes of galaxies at high redshifts are smaller than those of galaxies with a similar mass in the local Universe (see, e.g., Haghi
et al. (2015) and references therein).

There exist a multitude of scenarios to explain the physical processes of the evolution of galaxies that reproduce well their observed properties. There are a number of papers (G\'omez et al. 2010; Correa et al. 2015; Haghi et al. 2015; Sanders et al. 2020;
Armstrong et al. 2021) where various time-dependent
model gravitational potentials based on cosmological
models of the Universe were used to study the orbital
history of Galactic objects. Only a change in the
masses of the Galactic potential components are
considered in some papers (see, e.g., Armstrong
et al. 2021); both a change in the masses and a change in the sizes of the components are considered in other papers (G\'omez et al. 2010; Haghi et al. 2015).

In this paper we adopt the algorithm for constructing a semicosmological evolving model Galactic potential described in detail in G\'omez et al. (2010) and
Haghi et al. (2015). As the static potential or the
current potential, we adopt an axisymmetric threecomponent
model potential consisting of a spherical
bulge, a flat Miyamoto--Nagai disk (Miyamoto and
Nagai 1975), and a spherical Navarro--Frenk--White
dark matter halo (Navarro et al. 1997) modified by
us previously (Bajkova and Bobylev 2016) using the rotation curve from Bhattacharjee et al. (2014) in a wide range of Galactocentric distances (from 0 to 200 kpc).

The goal of our paper is to study the orbital history of 152 GCs from the list of Vasiliev (2019) based on up-to-date astrometric data (Vasiliev and Baumgardt
2021; Baumgardt and Vasiliev 2021) by numerically
integrating the orbits backward in time on
cosmological time scales comparable to the age of
the Universe. To perform a comparative analysis, we
use both static and evolving Galactic potentials. We
also set the goal to study the influence of separately a change in the masses and a change in the sizes of the potential components on the orbital motion of GCs.

The paper is structured as follows. Section 1 is devoted to the Galactic potential: we describe the static (Sect. 1.1) and evolving (Sect. 1.2) potentials
and provide basic relations for integrating the orbits
(Sect. 1.3). The data on 152 GCs are described
in Section 2. Section 3 is devoted to studying the
influence of a change in the masses and a change in
the sizes of the potential components on the orbital
motion of GCs. Section 4 presents the results of
our integration of the GC orbits in various potentials
(Sect. 4.1) and on various time scales (Sect. 4.2) and their comparative analysis. In the Conclusions we summarize our main results.

 \section{GALACTIC POTENTIAL}
 \subsection{Static potential}
As the Galaxy’s static gravitational potential we consider an axisymmetric potential consisting of three components: a central spherical bulge
$\Phi_b(r(R,Z))$, a disk $\Phi_b(r(R,Z))$, and a spherical dark
matter halo $\Phi_h(r(R,Z))$:
\begin{equation}
\begin{array}{lll}
  \Phi(R,Z)=\Phi_b(r(R,Z))+\Phi_d(r(R,Z))+\Phi_h(r(R,Z)).
 \end{array}
 \end{equation}
The potentials of the bulge $\Phi_b(r(R,Z))$ and the disk $\Phi_d(r(R,Z))$ are expressed in the form proposed by Miyamoto and Nagai (1975):
 \begin{equation}
  \Phi_b(r)=-\frac{M_b}{(r^2+a_b^2)^{1/2}},
  \label{bulge}
 \end{equation}
 \begin{equation}
 \Phi_d(R,Z)=-\frac{M_d}{\Biggl[R^2+\Bigl(a_d+\sqrt{Z^2+b_d^2}\Bigr)^2\Biggr]^{1/2}}.
 \label{disk}
\end{equation}
Here, we use a cylindrical coordinate system ($R,\psi,Z$) with the origin at the Galactic center. In a rectangular Cartesian coordinate system $(X,Y,Z)$ with the origin
at the Galactic center the distance to a star (spherical radius) is $r^2=X^2+Y^2+Z^2=R^2+Z^2$; $M_b$ and $M_d$ are the bulge and disk masses; $a_b, a_d,$ and $b_d$ are the scale lengths characterizing the sizes of the components.

In accordance with the convention adopted in Allen and Santillan (1991), in this paper the gravitational potential of all components is expressed in units of 100 km$^2$ s$^{-2}$, the distances are in kpc, and the masses are in units of the Galactic mass $M_{gal}=2.325\times10^7~M_\odot$ corresponding to the Gravitational constant $G=1.$

To describe the halo component, we used an expression in the Navarro--Frenk--White (NFW) form, which was used, in particular, in Irrgang et al. (2013)
and Bajkova and Bobylev (2016) to fit the model gravitational
potential to the latest data on the circular velocities of Galactic objects:
 \begin{equation}
  \Phi_h(r)=-\frac{M_h}{r} \ln {\Biggl(1+\frac{r}{a_h}\Biggr)},
 \end{equation}
Here, the weighting factor $M_h$ is equivalent to the mass inside a sphere of radius $\sim 5.3$ times the radial scale length $a_h$ (Irrgang et al. 2013).

Expression (4) for the halo potential is derived by jointly using the expression for the mass density from Navarro et al. (1997) (Eq. (1)) and the Poisson equation relating the mass density and the potential (see Irrgang et al. 2013; Bajkova and Bobylev 2016; Eq. (2)).

The parameters of the model Galactic potential adopted by us were derived in Bajkova and
Bobylev (2016) by fitting them to the data on the circular velocities of HI clouds, masers, and various halo objects with large Galactocentric distances $R$ (up to $\sim$200 kpc) from Bhattacharjee et al. (2014). To construct the rotation curve, we used $R_\odot=8.3$~kpc for the Galactocentric distance of the Sun and $V_\odot=244$ km s$^{-1}$ for the linear rotation velocity of the local standard of rest around the Galactic center. In addition, to fit the parameters, we used the constraints on the local dynamical mass density 
 $\rho_\odot=0.1 M_\odot$~pc$^{-3}$ and the force acting perpendicularly to the Galactic plane $|K_{Z=1.1}|/2\pi G=77M_\odot$~pc$^{-2}$ (Irrgang et al. 2013).

The parameters of the adopted static model are given in the first row of Table 1, where the masses of the components are expressed in units of the solar mass and the scale lengths are in kpc. The corresponding rotation curve is indicated by the red color in Fig. 1. This rotation curve corresponds to both the static potential and the variable one at present, i.e., at $z=0.$ According to this model (Bajkova and Bobylev 2016), the mass of the Galaxy is $M_{G_{(R \leq 200 kpc)}}=(0.75\pm0.19)\tilde{}\times10^{12}M_\odot$. This value agrees well with its present-day independent estimates. For example, the NFW halo mass estimated
quite recently by Koppelman and Helmi (2020) from the velocities of runaway halo stars is $M_{G_{(R \leq 200 kpc)}}=0.67^{+0.30}_{-0.15}\times10^{12}M_\odot$. The model gravitational potential of the Milky Way adopted by us seems more realistic than other known static model potentials, because it is supported by the data at large Galactocentric distances, which is very important in integrating the orbits of distant GCs and those with a large apocenter distance, and provides good agreement with the present-day estimates of the local parameters
and a number of independent Galactic mass estimates whose careful review is given in the recent paper by Wang et al. (2020). Note, however, that the peak in the innermost region (1--2 kpc), which is noticeable for the red and violet curves in Fig. 1,
may be slightly overestimated due to the potential at
the center being highly asymmetric because of the bar
(see Section 6.4.3 in the review of Bland-Hawthorn
and Gerhard (2016)). This problem requires a special
study, because there is no common view on it in
the literature. However, in our problem the rotation
curve at such small Galactocentric distances plays no
major role, because, as will be shown below, precisely
the orbits of objects far from the Galactic center are
subject to the greatest influence because of to the
evolution of the Galactic potential. Therefore, when solving our problem, we adhere to the traditional approaches to interpreting the rotation curve at small Galactocentric distances.

 \begin{table}
 \begin{center}
 \caption[]
 {\small\baselineskip=1.0ex
 Parameters of the model potential at present, 5, and 12 Gyr ago}
 \bigskip
 \begin{small}
 \label{t:model}
 \begin{tabular}{|l|l|l|l|l|}\hline
  Time $t$ &  Disk                   & Balge                   & Halo           \\\hline
    0 Gyr  & $M_d=6.51\times10^{10}$ & $M_b=1.03\times10^{10}$ & $M_h=2.90\times10^{11}$\\
  $z=0$    & $a_d=4.40; b_d=0.3084$  & $a_b=0.2672 $           & $a_h=7.70$    \\\hline
 $-$5 Gyr  & $M_d=4.70\times10^{10}$ & $M_b=7.45\times10^{9}$  & $M_h= 2.10\times10^{11}$\\
 $z=0.474$ & $a_d=3.02; b_d=0.2114$  & $a_b=0.1832$            & $a_h=5.28$    \\\hline
 $-$12 Gyr & $M_d=6.04\times10^{9}$  & $M_b=9.56\times10^{8}$  & $M_h= 2.69\times10^{10}$\\
 $z=3.426$ & $a_d=0.55; b_d=0.0383$  & $a_b=0.0331$            & $a_h=0.96$    \\\hline
 \end{tabular}
 \end{small}
 \end{center}
 \end{table}

\begin{figure*}
\begin{center}
   \includegraphics[width=0.6\textwidth]{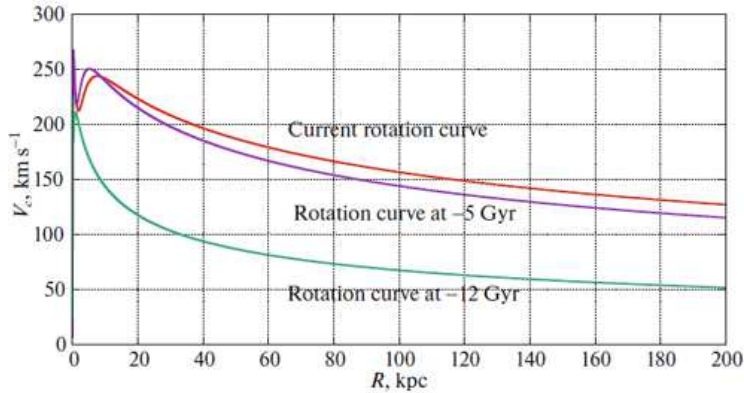}
  \caption{\small Galactic rotation curve at three epochs: the present time (red line), 5 Gyr ago (violet line), and 12 Gyr ago (green line).}
\label{f-rot}
\end{center}
\end{figure*}

 \subsection{Evolving Potential}
To construct an evolving Galactic potential, we adopt a semicosmologicalmodel in which the characteristic parameters determining the masses and sizes of the Galactic components change with time. We used the principle of constructing an evolving potential
considered in G\'omez et al.(2010) and Haghi
et al. (2015) (see also references in these papers).
However, our formulas slightly differ from those given
in these papers, because the expressions for the halo
potential differ. Our halo potential is specified by
Eq. (4), the parameters $M_h$ and $a_h,$ while in the above references the halo potential is specified via the virial mass, the virial radius, and the concentration parameter.

As a result, the algorithm for constructing an evolving potential adapted to our parameters, which retained the principles outlined in the papers cited above, looks as follows.

The evolution of the halo mass (4) as a function of redshift $z$ is specified by the expression
\begin{equation}
  M_h(z)=M_h(z=0)\exp(-2a_c z),
 \label{Mhz}
 \end{equation}
where the constant $a_c=0.34$ is defined as the halo formation epoch (G\'omez et al. 2010).

The following relation proposed by Bullock and Johnston (2005) is used for the disk and halo masses:
\begin{equation}
M_{d,b}(z) = M_h(z)\frac{M_{d,b}(z=0)}{M_h(z=0)},
\end{equation}
similarly, for the scale lengths of the components:
\begin{equation}
\{a_b, a_d, b_d\}(z)= a_h(z)\frac{\{a_b, a_d, b_d\}(z=0)}{a_h(z=0)},
\end{equation}
where the halo scale length $a_h(z)$ is calculated as
\begin{equation}
a_h(z) = \frac{K(z=0)a_h(z=0)}{K(z=0)},
\label{ah}
\end{equation}
\begin{equation}
  K(z) =\left( \frac{3M_h(z)}{4\pi\Delta_h(z)\rho_c(z)}\right) ^{1/3},
\label{K}
\end{equation}
where
\begin{equation}
\Delta_h(z)=18\pi^2+82[\Omega(z)-1]-39[\Omega(z)-1]^2.
\end{equation}
Here, $\Omega(z)$ is the mass density of the Universe,
\begin{equation}
\Omega(z) = \frac{\Omega_m (1+z)^3}{\Omega_m (1+z)^3+\Omega_\Lambda},
\end{equation}
and $\rho_c(z)$ is the critical density of the Universe at a given $z,$
\begin{equation}
\rho_c(z) = \frac{3H^2(z)}{8\pi G},
\end{equation}
where
\begin{equation}
H(z) = H_0 \sqrt{\Omega_\Lambda + \Omega_m (1+z)^3}.
\end{equation}
It is also assumed that the Universe is flat, in which the relation $\Omega_m+\Omega_\Lambda=1$ holds. We adopt the parameters $\Omega_m=0.3$ and $\Omega_\Lambda=0.7$. We take the Hubble constant in accordance with the result of the Planck mission, $H_0=68$ km s$^{-1}$ Mpc$^{-1}$ (Aghanim et al. 2020).

The relation between the redshift $z$ and the time $T$ elapsed since the beginning of the Big Bang looks as follows:
\begin{equation}
 z=\left(\frac{\Omega_m\sinh^2(\frac{3}{2}H_0 T\sqrt{\Omega_\Lambda})}{\Omega_\Lambda}\right)^{-1/3}-1.
 \label{H0}
\end{equation}
By setting $z=0,$ from Eq. (14) it is easy to derive the dependence of the product of the Hubble constant $H_0$ and the age of the Universe $T_0$ for the model of
a flat Universe on the parameters $\Omega_m$ and $\Omega_\Lambda$ ($\Omega_m + \Omega_\Lambda = 1$). In our case ($\Omega_m=0.3$ and $\Omega_\Lambda=0.7$), the
product $H_0\times T_0=0.9641$. Thus, at the Hubble constant $H_0=68$ km s$^{-1}$ Mpc$^{-1}$ the age of the Universe is $T_0=13.87$~Gyr.

\begin{figure*}
\begin{center}
   \includegraphics[width=0.8\textwidth]{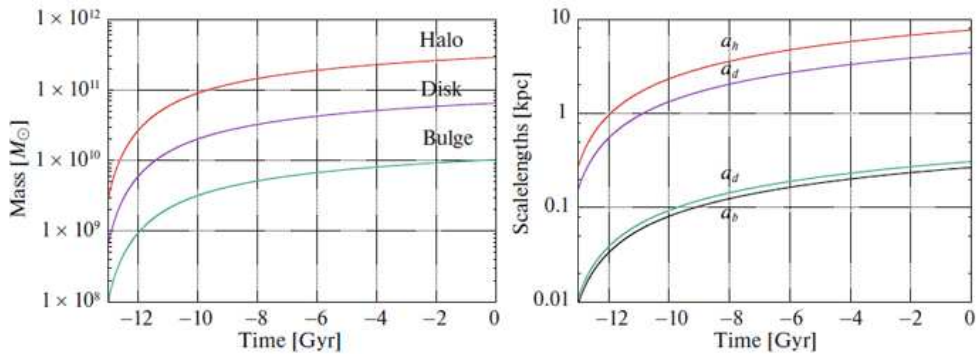}
  \caption{\small Time dependence of the masses (a) and scale lengths (b) of the Galactic components. A logarithmic scale is used.}
\label{f2}
\end{center}
\end{figure*}

Figure 2 presents the dependences of the masses and scale lengths of the Galactic components on cosmological time $t$ ($t=0$ corresponds to the present time, i.e., $z=0$). Note that the pattern of change in the masses and scale lengths of the components corresponds well to the dependences derived in Haghi et al. (2015). The parameters of the Galactic components
5 ($z=0.474$) and 12 ($z=3.426$) Gyr ago are given in the second and third rows of Table 1. The corresponding rotation curves are indicated in Fig. 1 by the violet and green colors. We presented the time dependences of the masses and scale lengths with a high accuracy in the form of eighth-degree power polynomials to be able to calculate the parameters for
any time $(t<0)$ when integrating the orbits.

 \subsection{Integration of Orbits}
The equation of motion for a test particle in an axisymmetric gravitational potential can be derived from the Lagrangian of the system $\pounds$ (see, e.g., Appendix A in Irrgang et al. (2013)):
\begin{equation}
 \begin{array}{lll}
 \pounds(R,Z,\dot{R},\dot{\psi},\dot{Z})=
 0.5(\dot{R}^2+(R\dot{\psi})^2+\dot{Z}^2)-\Phi(R,Z).
 \label{Lagr}
 \end{array}
\end{equation}
By introducing the canonical moments
\begin{equation}
 \begin{array}{lll}
    p_{R}=\partial\pounds/\partial\dot{R}=\dot{R},\\
 p_{\psi}=\partial\pounds/\partial\dot{\phi}=R^2\dot{\psi},\\
    p_{Z}=\partial\pounds/\partial\dot{Z}=\dot{Z},
 \label{moments}
 \end{array}
\end{equation}
we derive the Lagrange equations in the form of a
system of six first-order differential equations:
 \begin{equation}
 \begin{array}{llllll}
 \dot{R}=p_R,\\
 \dot{\psi}=p_{\psi}/R^2,\\
 \dot{Z}=p_Z,\\
 \dot{p_R}=-\partial\Phi(R,Z)/\partial R +p_{\psi}^2/R^3,\\
 \dot{p_{\psi}}=0,\\
 \dot{p_Z}=-\partial\Phi(R,Z)/\partial Z.
 \label{eq-motion}
 \end{array}
\end{equation}
To integrate Eqs. (17), we used the fourth-order Runge–Kutta algorithm.

The Sun’s peculiar velocity relative to the local standard of rest was taken to be $(u_\odot,v_\odot,w_\odot)=(11.1,12.2,7.3)\pm(0.7,0.5,0.4)$ km s$^{-1}$ (Sch\"onrich
et al. 2010). Here, we use the heliocentric velocities
in a moving Cartesian coordinate system with the
velocity $u$ directed toward the Galactic center, $v$ in the
direction of Galactic rotation, and $w$ perpendicular to
the Galactic plane and directed to the Galactic north pole.

Let the initial positions and space velocities of the test particle in the heliocentric coordinate system be $(x_o,y_o,z_o,u_o,v_o,w_o)$. Then, the initial positions $(X,Y,Z)$ and velocities $(U, V, W)$ of the test particle in Cartesian Galactic coordinates are specified by the formulas
\begin{equation}
 \begin{array}{llllll}
 X=R_\odot-x_o, Y=y_o, Z=z_o+h_\odot,\\
 R=\sqrt{X^2+Y^2},\\
 U=u_o+u_\odot,\\
 V=v_o+v_\odot+V_0,\\
 W=w_o+w_\odot,
 \label{init}
 \end{array}
\end{equation}
where $R_\odot$ and $V_\odot$ are the Galactocentric distance and
the linear rotation velocity of the local standard of rest
around the Galactic center, $h_\odot=16$~pc (Bobylev and Bajkova 2016) is the Sun’s height above the Galactic plane.

The initial radial, $\Pi$ ($\dot{R}$), and circular, $\Theta$ ($\dot{\psi}$), velocities
are specified by the expressions $\Pi=-U \frac{X}{R}+V \frac{Y}{R}$ and $\Theta=U \frac{Y}{R}+V \frac{X}{R}$, respectively.

\begin{figure*}
\begin{center}
   \includegraphics[width=0.5\textwidth]{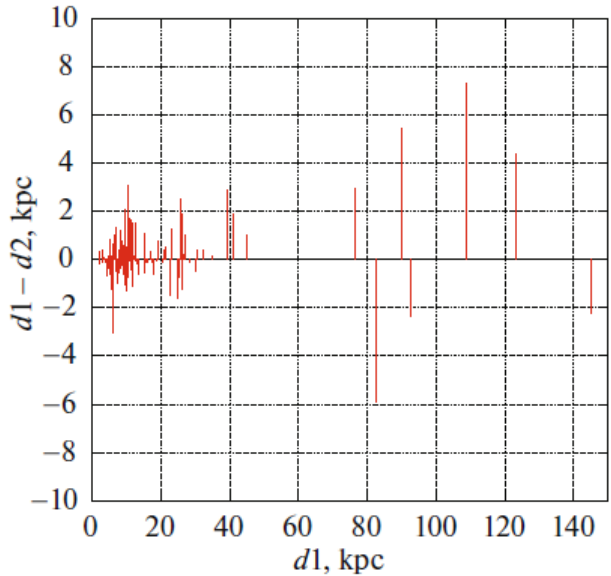}
\caption{\small Differences of the distances $(d1)$ ofGCs from the catalog by Harris (2010) and the mean distances $(d2)$ from Baumgardt and Vasiliev (2021) as a function of $d1.$}
 \label{f3}
\end{center}
\end{figure*}

 \section{DATA}
For the 152 GCs studied by us previously, whose orbits were published in Bajkova and Bobylev (2021a) based on the catalogue of Vasiliev (2019), we took the new mean proper motions and their uncertainties from the new catalogue by Vasiliev and Baumgardt (2021) produced from the data of the Gaia EDR3 catalogue. Figure 2 from Bajkova and Bobylev (2021b) compares the mean proper motions
from these two catalogues derived from Gaia DR2
and Gaia EDR3 measurements. As follows from
this figure, the new proper motions for a number of
GCs differ noticeably from the old ones. At the same
time, the measurement accuracy of the new proper motions, on average, doubled.

We used the new mean heliocentric distances from the recent paper by Baumgardt and Vasiliev (2021), whose accuracy is considerably higher than that of
the distances from the catalog by Harris (2010) that
were used previously in Bajkova and Bobylev (2021a,
2021b). The differences of the distances $d1$ of GCs
from the catalog by Harris (2010) and the mean distances
$d2$ from Baumgardt and Vasiliev (2021) as a
function of $d1$ are shown in Fig. 3. We see that the
distances changed quite significantly, especially for distant GCs.

All of the remaining astrometric data (the line-ofsight
velocities and coordinates) remained as before,
as in the catalogue of Vasiliev (2019).

\begin{figure*}
\begin{center}
   \includegraphics[width=0.9\textwidth]{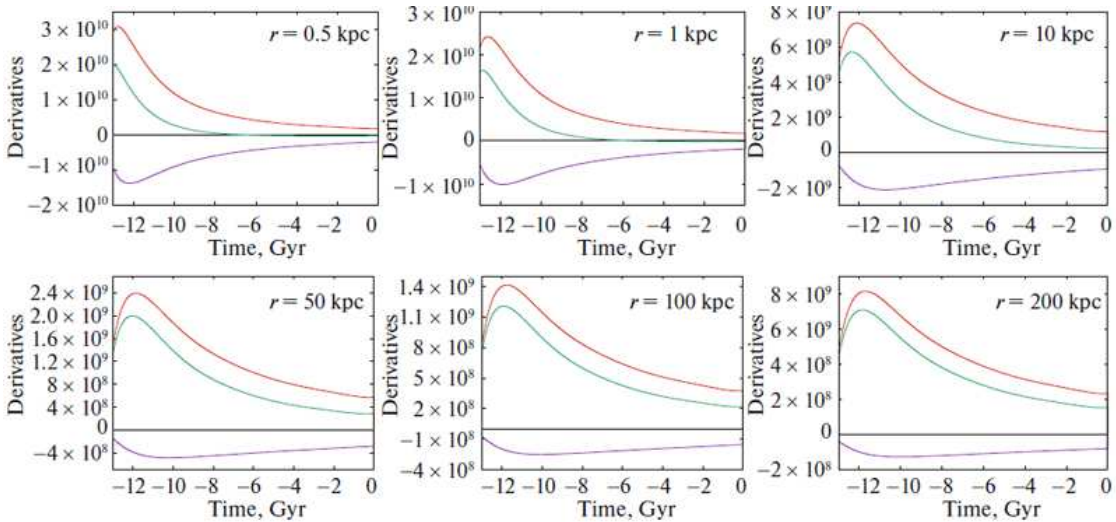}
\caption{\small Time derivatives of the potential of a spherical halo for various Galactocentric distances $r.$ The red and violet colors indicate the graphs of the derivative determined by the changing halo mass and the halo scale length (size), respectively; the green color indicates the graph of the total derivative.}
\label{f4}
\end{center}
\end{figure*}

 \section{STUDY OF THE EVOLVING POTENTIAL}
From Eqs. (2)--(4) for the potentials of the Galactic components one might expect the changes in the masses and scale lengths with time to act on the change in the potential in the opposite way. Indeed, the masses are in the numerator of the expressions
for the potential, while the scale lengths are in the
denominator. We studied separately the influence of
a change in the masses and the influence of a change in the scale lengths.

Consider this problem using the halo as an example. To determine the degree of influence of the parameters changing with time, let us find the time
derivatives of the potential:
\begin{equation}
\dot{\Phi}_h(t)= - \frac{\dot{M}_h(t)}{r}\ln \left ( 1+\frac{r}{a_h(t)}\right ) + \frac{M_h(t)\dot{a}_h(t)}{a_h(t)(a_h(t)+r)}.
\label{der}
\end{equation}
It follows from Eq. (19) that the first term is responsible for the rate of change of the halo potential because of the change in its mass $M_h(t)$, while the
second term is responsible for the rate of change of
the potential because of the change in the scale length
$a_h(t)$. Finding the derivatives of the functions $\dot{M}_h(t)$ and $\dot{a}_h(t)$ presents no problem, because they are fitted by power laws.

Figure 4 presents the time derivatives of the potential
of a spherical halo for various Galactocentric
distances $r.$ The components of the derivative determined
by the changing mass and scale length of
the halo are shown separately; the total derivative is
also given. The derivatives determined by the changing
mass and the changing scale length are seen to
have opposite signs. As follows from the graphs,
the maximum compensation of the influences occurs
at small radial distances $r.$ As the Galactocentric
distance increases, the relative influence of a change
in the mass grows compared to the influence of a
change in the scale length, although the absolute
values of all derivatives decrease with increasing $r.$
The maximum values of the derivatives are reached
in the time interval from $-13$ to $-11$~Gyr. Similar
dependences can be derived for the bulge and the disk.

To demonstrate the direct influence of a change in
the masses and scale lengths of the Galactic components
on the orbital motion of Galactic objects,
we simulated the motion of a test particle in static
and variable Galactic potentials in the time interval
from 0 to $-12$~Gyr. The results are presented in
Fig. 5. As a model orbit in the static potential we
chose a circular particle orbit with a radius of 8.3 kpc
indicated in the figure by the black color. The motion
of the test particle is shown for three forms of an
evolving potential, when (1) only the masses of the
components change, (2) only the scale lengths of the
components change, and (3) both masses and scale
lengths of the components change. As can be seen
from the figures, a change only in the masses of the
components causes an increase in the magnitude of
the particle radius vector during its motion backward
in time. On the contrary, a change only in the scale
lengths of the components causes it to decrease. In
the potential with a change in both masses and sizes
of the components, these two effects are added. The
figure (rightmost panel) also shows the change in the
magnitude of the particle radius vector with time for all of the potentials considered.

We would like to emphasize the importance of our study, because the evolution of the component sizes is sometimes ignored in the literature and only the evolution
of the mass is considered (see, e.g., Armstrong et al. 2021), which can lead to an overestimation of the change in the orbital parameters of Galactic objects.

\begin{figure*}
\begin{center}
   \includegraphics[width=0.9\textwidth]{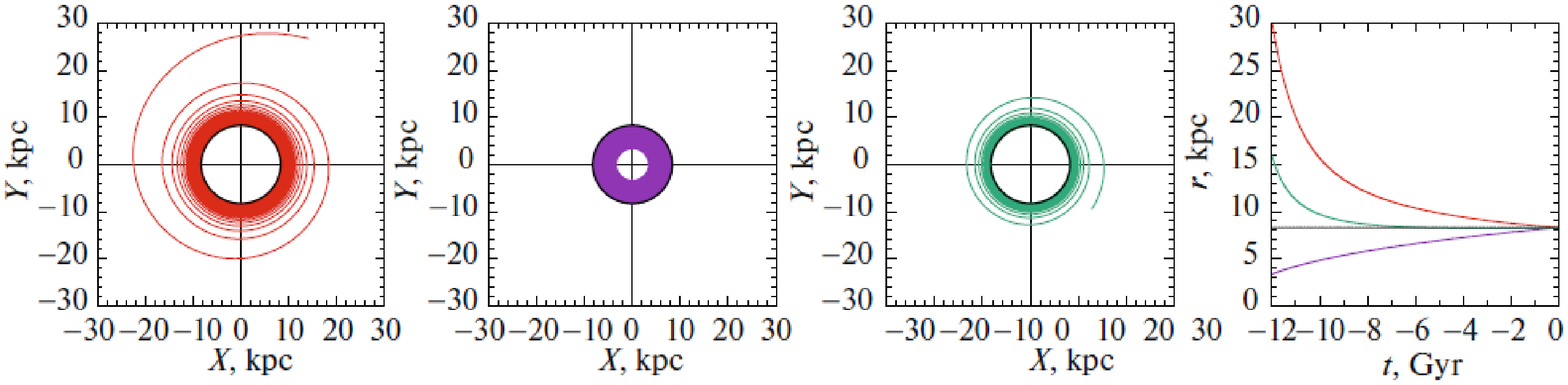}
\caption{\small Model particle orbits in three forms of an evolving potential in the time interval $[0,-12]$~Gyr: (1) only the masses of the Galactic components change (a), (2) only the sizes of the components change (b), and (3) both masses and sizes of the components change (c). For comparison, the black color on all panels indicates the circular orbit in a static potential. (d) The change in the particle radius vector with time for all forms of the potential.}
\label{f5}
\end{center}
\end{figure*}

 \section{COMPARISON OF THE ORBITAL PROPERTIES OF GALACTIC GLOBULAR CLUSTERS IN STATIC AND EVOLVING POTENTIALS}
 \subsection{Comparison of the Orbital Properties of GCs in a Potential only with Changing Masses and a Potential with Changing Masses and Sizes of the Components}
In this section we consider the orbital properties
of 152 Galactic GCs in the evolving potential constructed
by us in comparison with the static potential.

First we performed a comparative analysis of the
orbital motion of GCs when integrating the orbits
backward in time for 12~Gyr using three potentials:
(1) a static potential, (2) an evolving potential with
changing masses of the components, and (3) an
evolving potential with changing masses and changing
sizes of the Galactic components. Examples
of the orbits for six GCs (Eridanus, NGC 2419,
FSR 1758, NGS 104, NGC 362, Terzan 9) with different
Galactocentric distances belonging to different
MilkyWay subsystems (Massari et al. 2019; Bajkova
et al. 2020) are presented in Fig. 6. The orbits in
two projections, $(X,Y)$ and $(X,Z),$ are shown here.
In addition, the right panel shows the change in the
length of the radius vector $r$ of the orbit with time. The
orbits referring to different potentials are indicated by
different colors (see the caption to the figure).

\begin{figure*}
\begin{center}
   \includegraphics[width=0.7\textwidth]{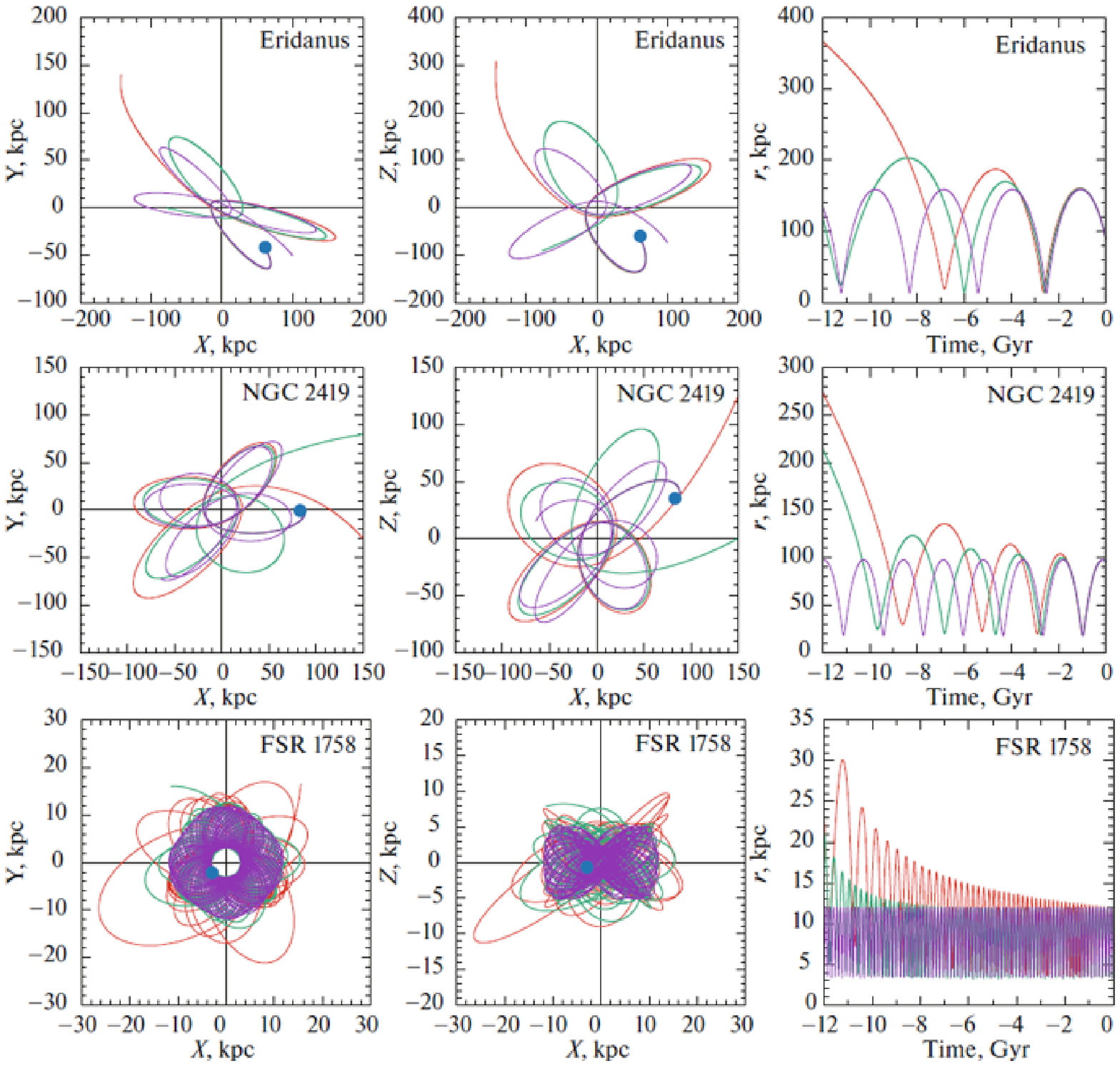}

   \includegraphics[width=0.7\textwidth]{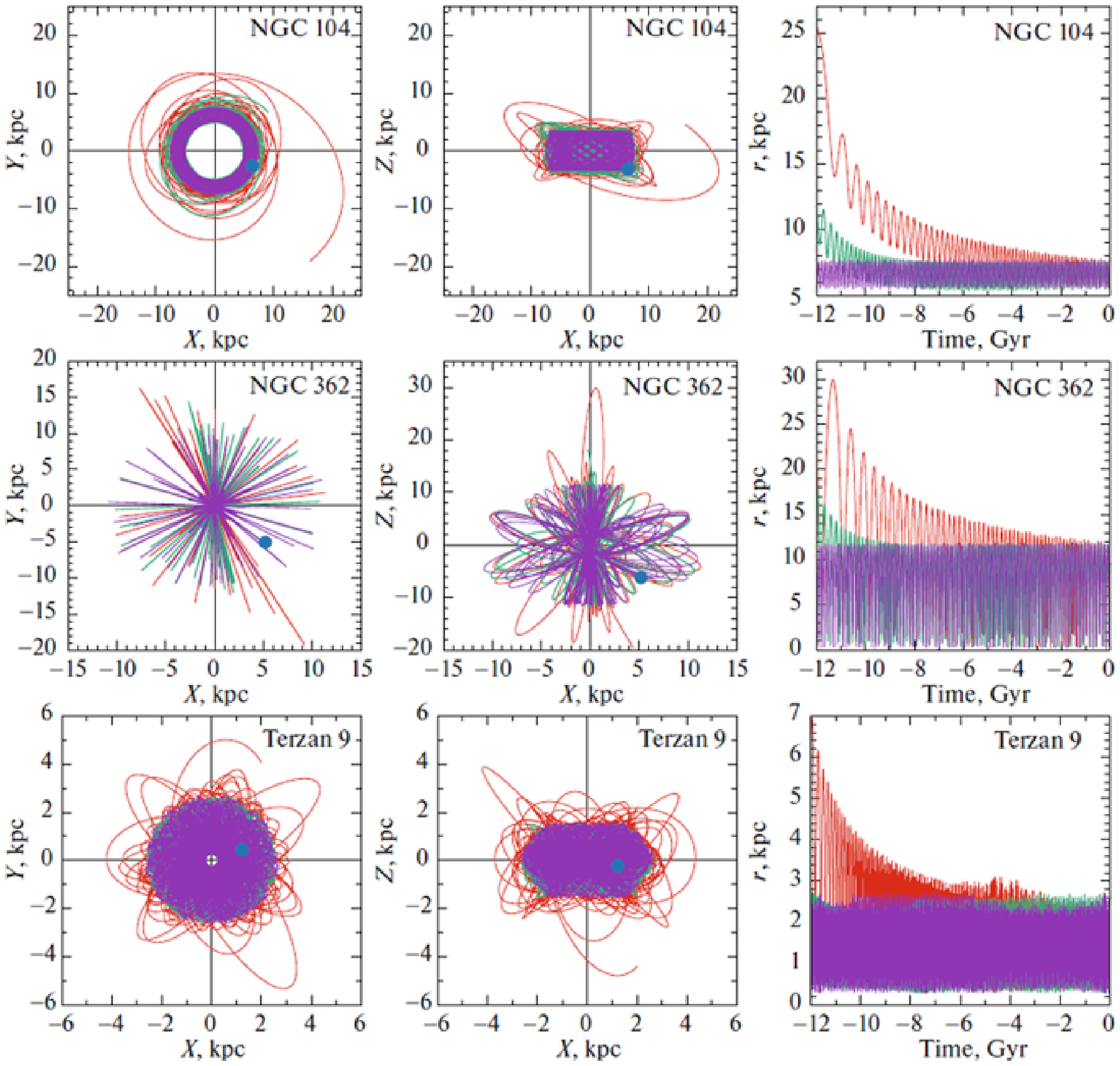}

\caption{\small Examples of the GC orbits in three potentials: (1) in a static potential, (2) in a potential with changing masses of the components, and (3) in a potential with changing masses and changing sizes of the components indicated by the violet, red,
and green colors, respectively. The orbits were integrated for 12 Gyr backward. The beginning of the orbits is marked by the blue circle.}
\label{f-6-I}
\end{center}
\end{figure*}


As follows from Fig.~6, compared to the orbits in the static potential, the orbits in the potential only with evolving masses of the components underwent the biggest change. In the potential with evolving masses and sizes of the components the change in
the sizes compensated significantly for the effect from
the change in the masses; the closer the object to the Galactic center, the stronger the compensation. Using Terzan~9 closest (of all the GCs considered) to the Galactic center as an example, we can see an almost complete overlap of the orbits in the static and
evolving potentials.

Figure 7 shows the distribution of GCs in the
$(X,Y)$ and $(X,Z)$ planes 13 Gyr ago in the static
potential, the potential only with changing masses
of the components, and the potential with changing
masses and changing sizes of the components. The
region of the distribution of GCs in the potential with
minimum masses and maximum invariable sizes of
the components is seen to be biggest. In the potential
with minimum masses and minimum sizes of
the components the distribution occupies a smaller
region. The minimum size of the distribution is observed
in the static potential, when the masses of the
Galactic components are maximal.

Such orbital parameters as the apocenter distance $(apo),$ the pericenter distance $(peri),$ the maximum elevation in $Z (Z_{max}),$ and the eccentricity $(ecc)$ in the static potential are compared with the parameters in the potential only with changing masses of the
components and the potential with changing masses
and changing sizes of the components in Fig. 8. It
can be seen from the figures that the pericenter distances
of GCs were subject to minimum change. The remaining parameters changed rather significantly, especially in the potential only with changing masses
of the components. In the potential with changing masses and sizes of the components the change in the parameters was smaller, which is to be expected in accordance with the results of our study performed in the previous section.

\begin{figure*}
\begin{center}
   \includegraphics[width=0.8\textwidth]{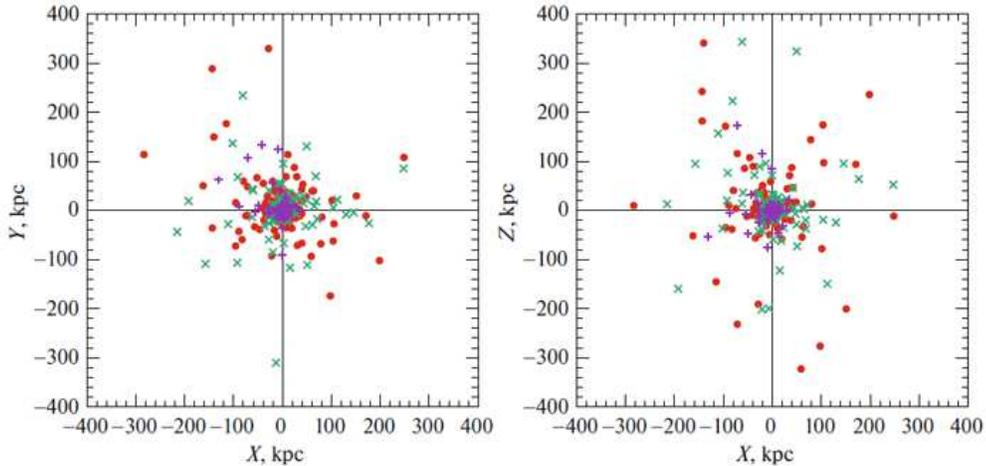}
\caption{\small Distribution of GCs in the $(X,Y)$ and $(X,Z)$ planes 13 Gyr ago in the static potential (violet crosses), the potential only with changing masses of the components (red circles), and the potential with changing masses and changing sizes of the
components (green crosses).}
\label{f-7}
\end{center}
\end{figure*}

\begin{figure*}
\begin{center}
   \includegraphics[width=0.8\textwidth]{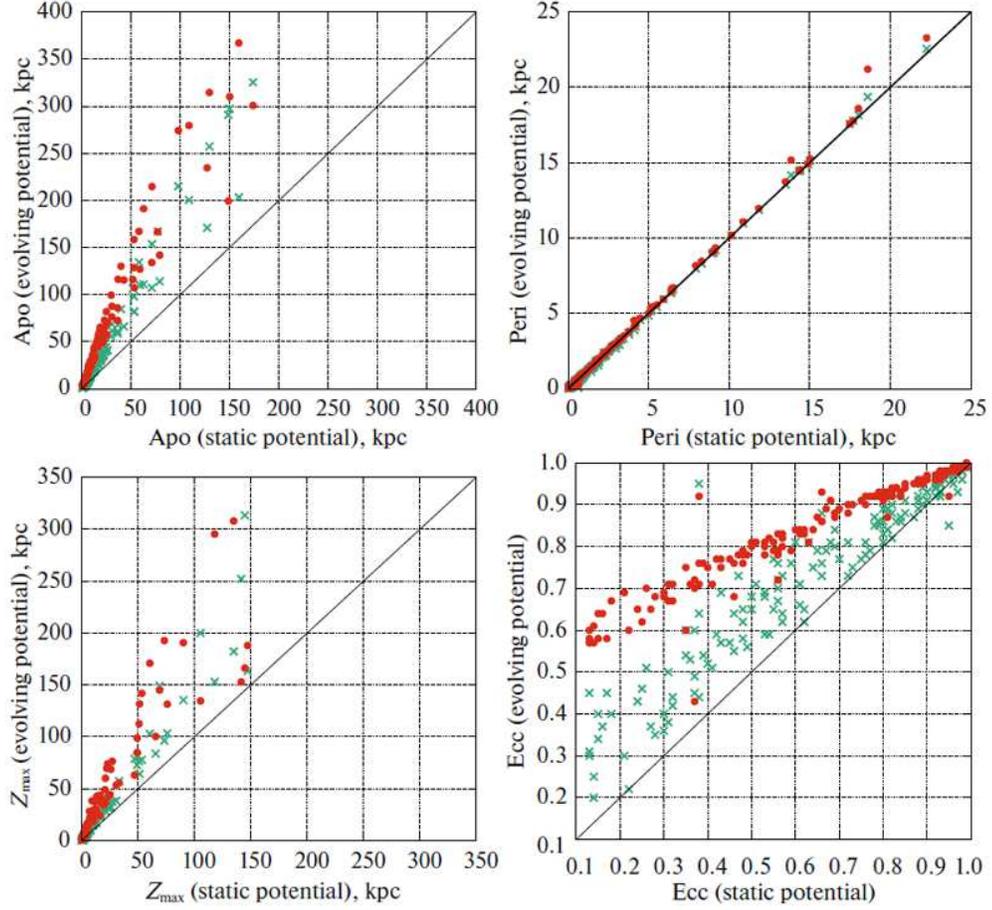}
\caption{\small Comparison of the orbital parameters $(apo, peri, Z_{max}, ecc)$ derived in the static potential with the parameters derived in the potential only with changing masses of the components (red circles) and the potential with changing masses and changing sizes of the components (green crosses). The orbits were integrated for 12 Gyr backward. The diagonal coincidence line is plotted on each panel.}
\label{f-8}
\end{center}
\end{figure*}

 \subsection{Comparison of the Orbital Properties of GCs in a Potential with Changing Masses and Sizes of the Components over the Periods $[0,-5]$ and $[0,-12]$ Gyr}
In this section we present the results of our integration
of the GC orbits in an evolving potential
with changing masses and sizes of the components
in the time intervals $[0,-5]$ and $[0,-12]$~Gyr. The
parameters of the potential components at $-5$ and
$-12$~Gyr are given in Table 1. The corresponding
rotation curves are shown in Fig. 1.

The orbital parameters $apo, peri, Z_{max},$ and $ecc$ of
all 152 GCs in the static and evolving potentials in
the time intervals $[0,-5]$ and $[0,-12]$ Gyr are given in
Table~2. The orbital parameters in the evolving potential
are compared with those in the static potential in
Fig.~9. It follows from this figure that the discrepancy
between the parameter is significant in the time interval
$[0,-12]$ Gyr, while in the time interval $[0,-5]$~Gyr
the orbital parameters of GCs undergo, on average,
minor changes compared to the orbital parameters in
the static potential, which fit well into the limits of
the statistical uncertainties caused by the errors in the
data. Regarding the latter thesis, a similar conclusion
was also drawn in Sanders et al. (2020), where the
time evolution over the last 5~Gyr was shown to introduce
an uncertainty in the orbital parameters of the
Milky Way satellites comparable to the uncertainty
caused by the observational errors or the uncertainty
in the current Milky Way potential.

\begin{figure*}
\begin{center}
   \includegraphics[width=0.8\textwidth]{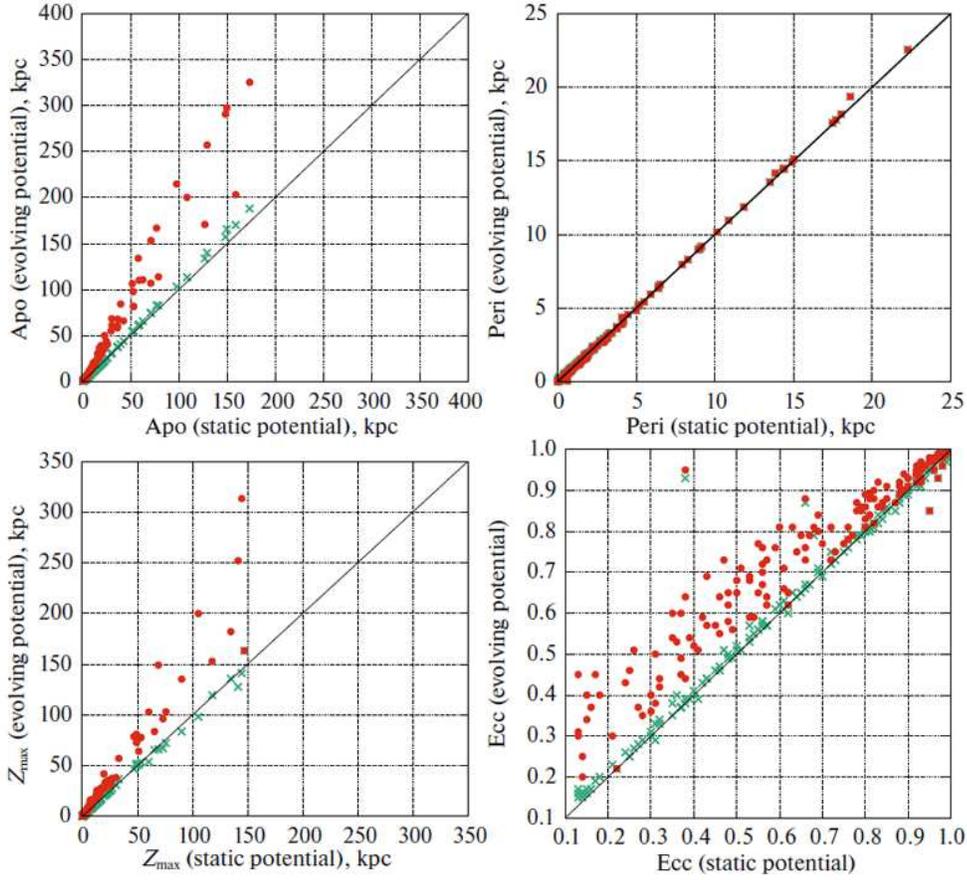}
\caption{\small Comparison of the orbital parameters $(apo, peri, Z_{max}, ecc)$ in the static potential with those in the potential with changing masses and changing scale lengths of the components over the periods $[0,-5]$ (green crosses) and $[0,-12]$ Gyr (red circles). The diagonal coincidence line is plotted on each panel. The rms deviations of the differences in relative units for $apo, peri, Z_{max}$, and $ecc$ are, respectively, 1.3, 1.9, 1.2, 5.2\% in the time interval $[0,-5]$ Gyr and 16, 1.9, 14.6, 13\% in the time
interval $[0,-12]$~Gyr.}
\label{f-9}
\end{center}
\end{figure*}

\begin{figure*}
\begin{center}
   \includegraphics[width=0.7\textwidth]{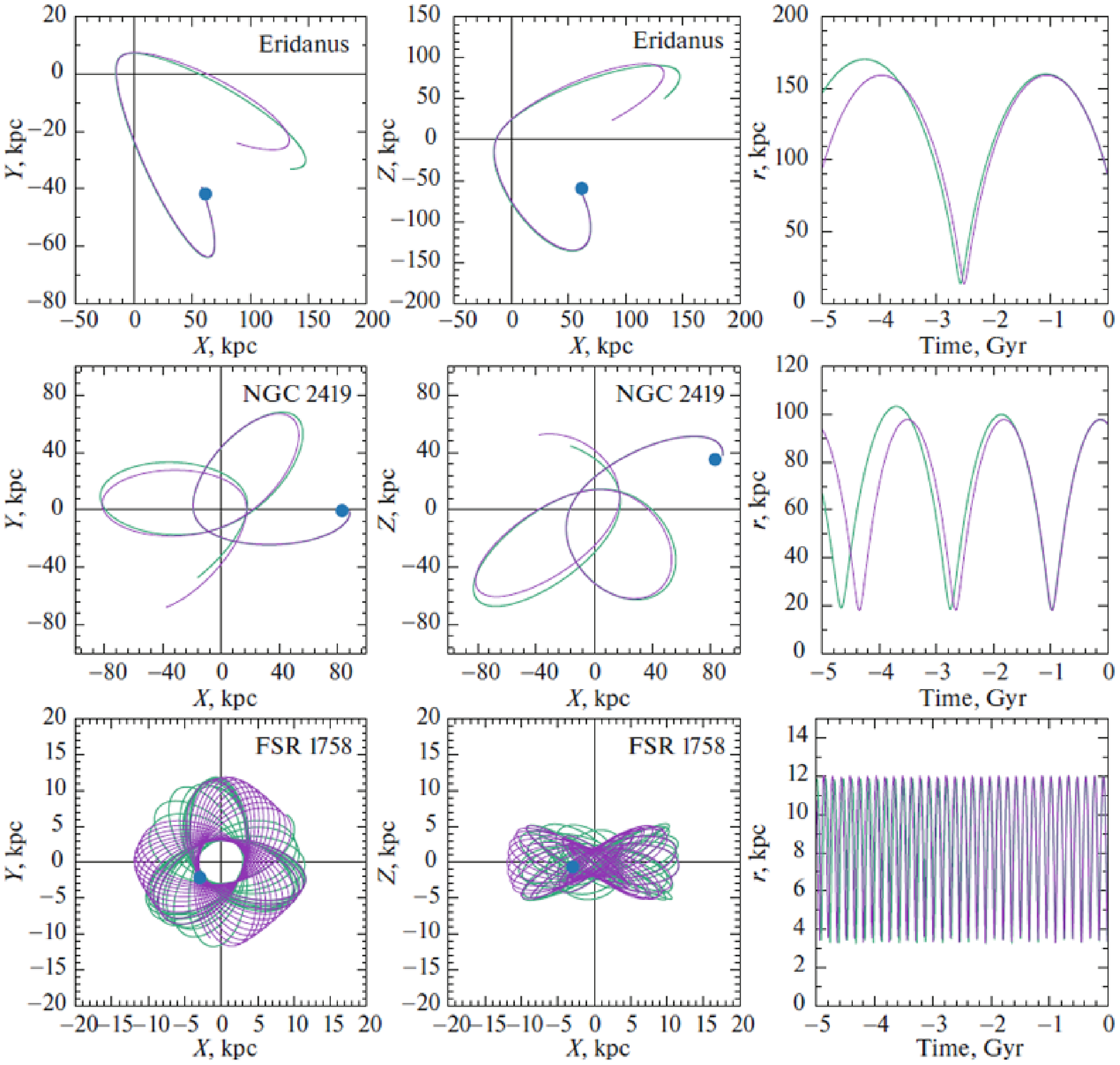}

   \includegraphics[width=0.7\textwidth]{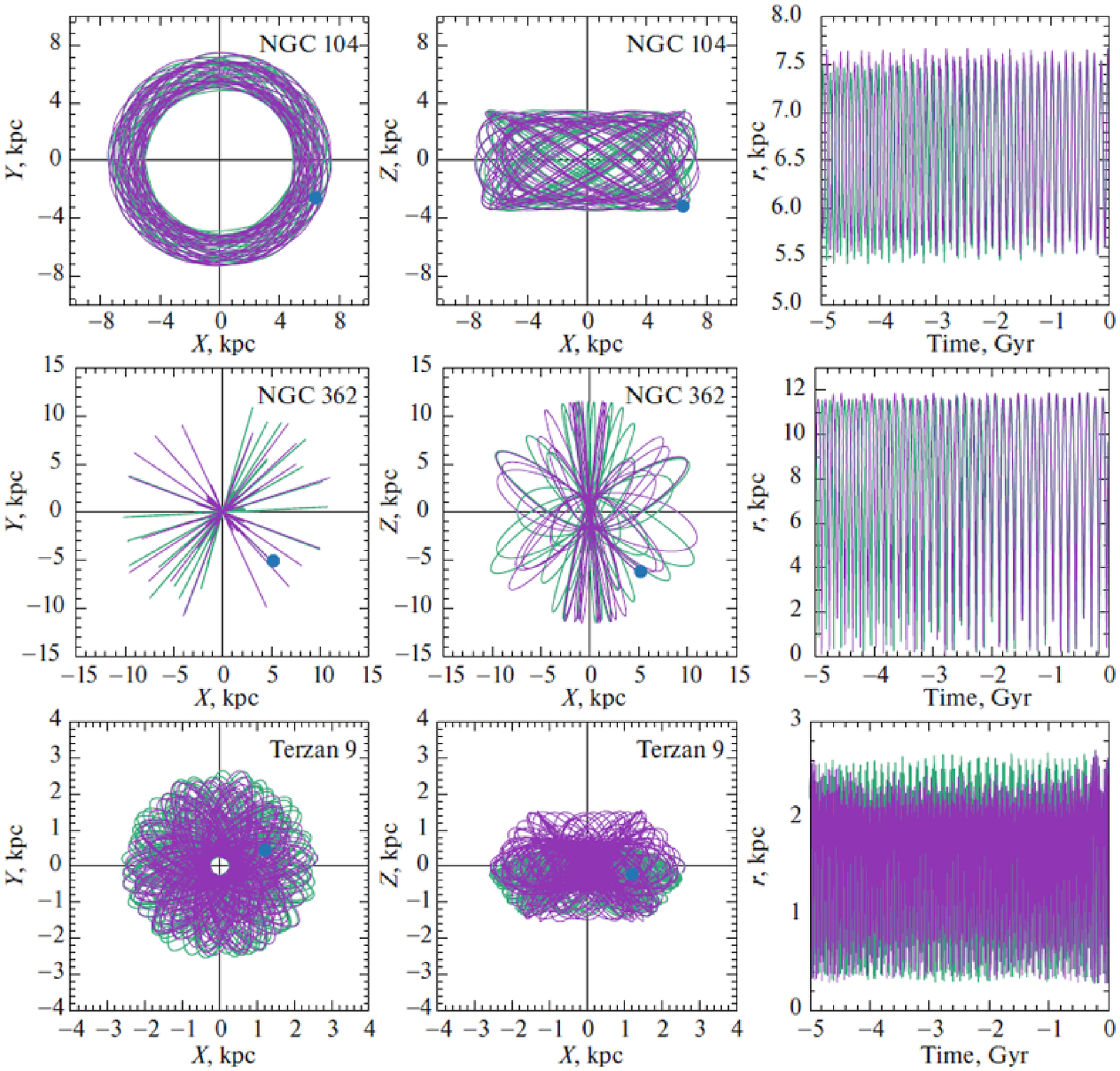}

\caption{\small Examples of the GC orbits in the static potential and the potential with changing masses and changing sizes of the components indicated by the violet and green colors, respectively. The orbits were integrated for 5 Gyr backward. The
beginning of the orbits is marked by the blue circle.}
\label{f-10}
\end{center}
\end{figure*}

As shown in Bajkova and Bobylev (2021b), the rms deviation of the uncertainty caused by the errors in the proper motions, line-of-sight velocities, and distances, which dominate, is 7.6\% in relative units (i.e., in the units calculated with respect to
the effective range of change in the parameter) for
the orbital eccentricity and 3--4\% for the remaining
parameters. The rms deviations of the differences in
relative units for the eccentricity, $Z_{max}, apo,$ and $peri$
when integrating the orbits in the evolving and static
potentials in the time interval $[0,-5]$ Gyr are 5.2,
1.2, 1.3, 1.9\%, respectively, while the corresponding
deviations when integrating the orbits in the interval
$[0,-12]$~Gyr are considerably larger (except for the
pericenter distance) or, more specifically, 13, 14.6, 16,
1.9\%.

As an illustration, Fig. 10 shows the orbits of the
same GCs as those in Fig. 6 in the static and evolving
potentials in the time interval $[0,-5]$~Gyr.

Thus, on time scales less than 5 Gyr the Galactic potential may be deemed constant and, in particular, the catalogue of Bajkova and Bobylev (2021a) may be used to get an idea of the orbital motion of Galactic GCs on these time scales.

We also calculated the infall time onto the Galaxy for each GC from the change of sign of the total energy from positive to negative, which is equivalent to the detection of the earliest time when the GC crosses the virial radius of its parent dark matter halo. Table 3 gives the GC infall times onto the Galaxy after $-13$~Gyr. The table also gives the
classification of GCs by Galactic subsystems (GS) (Massari et al. 2019; Bajkova and Bobylev 2021a) (the following designations are adopted: D---disk, GE---Sausage or Gaia--Enceladus galaxy, H99---Helmi stream, Seq---Sequoia galaxy, Sgr---Sagittarius dwarf galaxy, HE---unassociated high-energy group.
As follows from the table, there are 20 GE objects
out of 38 entering into the entire sample, seven Sgr
(out of seven), seven H99 (out of eight), six HE (out
of six), five D (out of 35), and four Seq (out of nine) in the list of GCs that fell onto the Galaxy later than $-13$~Gyr.

 \section*{CONCLUSIONS} 
Many authors still use an invariable, static potential to study the orbital motion of Galactic objects, despite the fact that a number of works (see, e.g.,
Armstrong et al. (2021), Haghi et al. (2015), and references therein) on the construction of an evolving potential based on the models describing the current Galactic potential and cosmological models of the Universe have appeared in recent years. In this paper,
to study the orbital motion of GCs, we constructed an evolving Galactic potential according to the algorithm
based on a semicosmological model and developed in detail in G\'omez et al. (2010) and Haghi et al. (2015). As the static model potential and the current potential we consider an axisymmetric three-component potential with a bulge and a disk in the Miyamoto--Nagai (1975) form and with a spherical Navarro--Frenk--White (1997) halo modified in Bajkova and Bobylev (2016) using the rotation curve from Bhattacharjee et al. (2014) with Galactocentric distances of objects up to 200 kpc.

Both the masses and the scale lengths determining
the sizes of the components change with time
in the constructed evolving potential. For the first
time we have studied separately the influence of a
change in the masses and a change in the sizes of the
Galactic components. We showed that the changes
in the masses and sizes of the components act in the
opposite way. At small Galactocentric distances this
influence is maximally compensated for. The orbits
of distant objects and objects with a large apocenter
distance experience the greatest influence. Using six
real GCs with various Galactocentric distances as
an example, we showed the effect from a change in
the masses and scale lengths of the potential components.
Our study is important in that the evolution
of the component sizes is sometimes ignored in the
literature and only the evolution of the masses is
considered, which may lead to an overestimation of
the change in the orbital history of Galactic objects.

\begin{table*}
 \begin{center}
 \caption[]
 {\small\baselineskip=1.0ex
 Orbital parameters of GCs in the static potential (the superscript ``st'') and the potential with changing masses and changing sizes of the components (the subscripts ``-5'' and ``-12'' refer to the periods $[0,-5]$ and $[0,-12]$ Gyr, respectively)}
  \bigskip
 \begin{scriptsize} 
 \label{t:param}
 \begin{tabular}{|l|r|r|r|r||r|r|r|r||r|r|r|r|}\hline
  Name & $apo^{st}$, &$peri^{st}$,&$ecc^{st}$&$Z_{\rm max}^{st}$, & $apo^{-5}$, &$peri^{-5}$,&$ecc^{-5}$&$Z_{\rm max}^{-5}$,&$apo^{-12}$, &$peri^{-12},$&$ecc^{-12}$&$Z_{\rm max}^{-12}$,\\
  &kpc &kpc  &   &kpc &kpc &kpc   &   &kpc&kpc&kpc&   &kpc\\\hline
NGC 104  &    7.7&   5.51& 0.16&    3.5&    7.7&   5.42& 0.17&    3.5&   11.9&   5.42& 0.37&    5.1\\
NGC 288  &   12.4&   1.43& 0.79&   10.2&   12.4&   1.46& 0.79&   10.3&   18.9&   1.46& 0.86&   16.3\\
NGC 362  &   11.9&   0.08& 0.99&   11.6&   11.9&   0.10& 0.98&   11.5&   18.2&   0.10& 0.99&   18.2\\
Whiting 1 &   79.0&  22.25& 0.56&   75.9&   82.8&  22.55& 0.57&   72.5&  114.0&  22.55& 0.67&  103.2\\
NGC 1261 &   21.3&   0.84& 0.92&   16.9&   21.4&   0.93& 0.92&   17.2&   33.9&   0.93& 0.95&   25.8\\
Pal 1    &   19.4&  14.89& 0.13&    4.8&   20.0&  14.89& 0.15&    5.0&   39.3&  14.89& 0.45&    8.6\\
E 1      &  129.4&   5.19& 0.92&  117.8&  140.1&   5.23& 0.93&  119.6&  257.2&   5.23& 0.96&  152.9\\
Eridanus &  159.2&  13.84& 0.84&  134.8&  170.3&  14.16& 0.85&  135.8&  203.2&  14.16& 0.87&  182.1\\
Pal 2    &   39.8&   1.35& 0.93&    9.1&   40.9&   1.31& 0.94&    8.5&   84.5&   1.31& 0.97&   16.3\\
NGC 1851 &   19.9&   0.12& 0.99&   16.8&   19.9&   0.15& 0.99&   18.1&   31.4&   0.15& 0.99&   23.8\\
NGC 1904 &   19.9&   0.24& 0.98&   14.7&   19.9&   0.26& 0.97&   14.2&   31.6&   0.26& 0.98&   25.9\\
NGC 2298 &   16.9&   0.49& 0.94&   12.0&   16.7&   0.47& 0.95&   12.0&   25.6&   0.47& 0.96&   17.3\\
NGC 2419 &   97.8&  18.02& 0.69&   73.2&  103.3&  18.16& 0.70&   67.0&  214.8&  18.16& 0.84&   96.2\\
Pyxis    &  173.6&  18.60& 0.81&  144.6&  188.1&  19.36& 0.81&  141.1&  325.2&  19.36& 0.89&  313.4\\
NGC 2808 &   14.9&   0.90& 0.89&    5.2&   14.8&   0.78& 0.90&    5.2&   22.9&   0.74& 0.94&    9.3\\
E 3      &   12.4&   9.05& 0.15&    5.4&   12.4&   9.05& 0.16&    5.5&   20.9&   9.05& 0.40&    8.7\\
Pal 3    &  148.5&  68.08& 0.37&  141.4&  157.4&  72.86& 0.37&  127.7&  290.7&  72.86& 0.60&  252.2\\
NGC 3201 &   24.9&   8.29& 0.50&   10.3&   25.6&   8.30& 0.51&   10.8&   44.1&   8.30& 0.68&   15.2\\
Pal 4    &  108.7&   4.10& 0.93&  105.3&  113.6&   4.37& 0.93&   98.3&  200.2&   4.37& 0.96&  199.9\\
Crater   &  149.9&  71.69& 0.35&  147.1&  165.7&  74.72& 0.38&  163.3&  297.3&  74.72& 0.60&  163.3\\
NGC 4147 &   25.5&   0.79& 0.94&   25.1&   25.7&   0.88& 0.93&   25.2&   40.0&   0.88& 0.96&   31.1\\
NGC 4372 &    7.3&   2.96& 0.42&    2.1&    7.3&   2.88& 0.43&    2.1&   11.2&   2.85& 0.59&    3.6\\
Rup 106  &   36.8&   4.48& 0.78&   22.9&   37.9&   4.56& 0.79&   23.2&   68.1&   4.56& 0.87&   31.6\\
NGC 4590 &   30.5&   8.94& 0.55&   19.5&   31.5&   8.96& 0.56&   20.1&   68.4&   8.96& 0.77&   42.0\\
NGC 4833 &    8.0&   0.63& 0.85&    3.7&    8.0&   0.60& 0.86&    3.6&   10.9&   0.52& 0.91&    5.6\\
NGC 5024 &   23.0&   9.15& 0.43&   22.2&   23.5&   9.19& 0.44&   22.4&   50.3&   9.19& 0.69&   33.0\\
NGC 5053 &   18.1&  10.87& 0.25&   17.5&   18.4&  10.96& 0.25&   17.5&   29.8&  10.96& 0.46&   28.3\\
NGC 5139 &    7.1&   1.28& 0.70&    3.0&    7.1&   1.28& 0.69&    3.0&   10.0&   1.28& 0.77&    5.4\\
NGC 5272 &   15.9&   5.14& 0.51&   13.3&   15.9&   5.15& 0.51&   13.5&   30.5&   5.15& 0.71&   23.5\\
NGC 5286 &   13.0&   0.54& 0.92&    7.5&   13.0&   0.56& 0.92&    8.9&   19.2&   0.56& 0.94&   16.4\\
NGC 5466 &   52.9&   5.92& 0.80&   49.2&   54.9&   5.94& 0.80&   49.9&   97.8&   5.94& 0.89&   72.8\\
NGC 5634 &   22.3&   2.29& 0.81&   20.6&   22.4&   2.36& 0.81&   21.2&   35.6&   2.36& 0.88&   32.5\\
NGC 5694 &   71.0&   2.75& 0.93&   49.1&   74.7&   2.80& 0.93&   52.0&  107.1&   2.80& 0.95&   80.9\\
IC 4499  &   29.9&   6.43& 0.65&   27.1&   30.6&   6.49& 0.65&   27.6&   55.2&   6.49& 0.79&   37.5\\
NGC 5824 &   36.4&  13.50& 0.46&   30.6&   38.0&  13.55& 0.47&   31.1&   60.7&  13.55& 0.64&   38.3\\
Pal 5    &   17.6&   7.93& 0.38&   16.2&   17.7&   7.96& 0.38&   16.2&   36.4&   7.96& 0.64&   22.3\\
NGC 5897 &    8.8&   1.94& 0.64&    7.6&    8.8&   1.95& 0.64&    7.6&   13.7&   1.95& 0.75&   10.9\\
NGC 5904 &   23.3&   2.23& 0.82&   21.0&   23.4&   2.39& 0.81&   21.8&   38.7&   2.39& 0.88&   28.1\\
NGC 5927 &    5.5&   4.13& 0.15&    0.8&    5.5&   3.94& 0.17&    0.8&    7.9&   3.89& 0.34&    1.6\\
NGC 5946 &    5.8&   0.06& 0.98&    4.3&    5.7&   0.06& 0.98&    4.3&    7.2&   0.06& 0.98&    4.3\\
ESO 224-8 &   16.8&  11.84& 0.17&    1.9&   17.3&  11.86& 0.19&    2.0&   30.9&  11.86& 0.45&    4.0\\
NGC 5986 &    5.6&   0.20& 0.93&    3.9&    5.1&   0.25& 0.91&    3.8&    6.3&   0.25& 0.92&    4.8\\
FSR 1716 &    5.2&   2.20& 0.41&    1.6&    5.0&   2.18& 0.39&    1.6&    6.6&   2.13& 0.51&    2.3\\
Pal 14   &  127.1&   1.49& 0.98&   90.1&  133.9&   1.49& 0.98&   83.7&  170.9&   1.49& 0.98&  135.3\\
BH 184   &    4.7&   1.65& 0.48&    1.5&    4.7&   1.53& 0.50&    1.5&    5.7&   1.52& 0.58&    2.2\\
NGC 6093 &    4.2&   0.45& 0.80&    3.9&    4.2&   0.45& 0.81&    3.9&    5.0&   0.45& 0.83&    4.7\\
NGC 6121 &    6.8&   0.61& 0.84&    3.1&    6.8&   0.62& 0.83&    2.9&    8.6&   0.62& 0.86&    4.4\\
NGC 6101 &   36.3&  10.14& 0.56&   21.4&   37.8&  10.16& 0.58&   21.5&   58.6&  10.16& 0.70&   34.2\\
NGC 6144 &    3.4&   1.56& 0.37&    3.2&    3.4&   1.55& 0.37&    3.2&    4.1&   1.55& 0.45&    3.9\\
NGC 6139 &    3.6&   0.97& 0.57&    2.7&    3.6&   0.97& 0.57&    2.7&    4.2&   0.97& 0.62&    3.0\\
\hline
  \end{tabular}
 \end{scriptsize}
  \end{center}
  \end{table*}

\begin{table*}
 \begin{center}
 \centerline{\small {\bf Table 2.} (Contd.)}
 \bigskip
 \begin{scriptsize}
 \label{t:par1}
 \begin{tabular}{|l|r|r|r|r||r|r|r|r||r|r|r|r|}\hline
  Name & $apo^{st},$ &$peri^{st}$,&$ecc^{st}$&$Z_{\rm max}^{st}$, & $apo^{-5}$, &$peri^{-5}$&$ecc^{-5}$&$Z_{\rm max}^{-5}$,&$apo^{-12}$, &$peri^{-12}$,&$ecc^{-12}$&$Z_{\rm max}^{-12}$,\\
  &kpc &kpc  &   &kpc &kpc &kpc   &   &kpc&kpc&kpc&   &kpc\\\hline
Terzan 3 &    3.1&   2.33& 0.14&    1.7&    3.1&   2.24& 0.16&    1.7&    3.6&   2.19& 0.25&    2.2\\
NGC 6171 &    4.0&   1.07& 0.57&    2.5&    3.9&   1.07& 0.57&    2.4&    4.8&   1.07& 0.64&    3.1\\
ESO 452-11 &    3.0&   0.06& 0.96&    2.2&    2.9&   0.06& 0.96&    2.1&    3.0&   0.05& 0.97&    2.4\\
NGC 6205 &    8.8&   0.97& 0.80&    7.8&    8.8&   0.97& 0.80&    7.8&   12.7&   0.97& 0.86&    9.6\\
NGC 6229 &   30.6&   0.57& 0.96&   23.9&   31.1&   0.60& 0.96&   24.5&   61.3&   0.60& 0.98&   35.9\\
NGC 6218 &    4.9&   2.08& 0.40&    2.8&    4.9&   2.05& 0.41&    2.8&    6.4&   2.04& 0.52&    4.1\\
FSR 1735 &    4.2&   0.21& 0.90&    2.9&    4.2&   0.20& 0.91&    2.6&    4.4&   0.20& 0.91&    3.6\\
NGC 6235 &    7.2&   3.13& 0.39&    4.9&    7.1&   3.12& 0.39&    4.9&   10.5&   3.12& 0.54&    6.8\\
NGC 6254 &    4.8&   1.78& 0.46&    2.8&    4.8&   1.76& 0.46&    2.8&    6.1&   1.76& 0.55&    4.1\\
NGC 6256 &    2.4&   1.53& 0.22&    0.7&    2.4&   1.52& 0.22&    0.7&    2.4&   1.52& 0.22&    0.9\\
Pal 15   &   52.9&   1.30& 0.95&   51.3&   55.1&   1.31& 0.95&   51.5&   82.0&   1.31& 0.97&   64.3\\
NGC 6266 &    2.7&   0.84& 0.53&    1.0&    2.7&   0.83& 0.53&    1.0&    2.9&   0.74& 0.59&    1.0\\
NGC 6273 &    3.5&   0.85& 0.61&    3.4&    3.5&   0.85& 0.61&    3.4&    4.2&   0.85& 0.66&    3.9\\
NGC 6284 &    6.4&   0.51& 0.85&    5.7&    6.4&   0.54& 0.85&    5.6&    8.6&   0.54& 0.88&    7.1\\
NGC 6287 &    4.4&   0.48& 0.81&    4.2&    4.4&   0.48& 0.80&    4.2&    5.4&   0.48& 0.84&    4.9\\
NGC 6293 &    3.2&   0.13& 0.92&    2.3&    3.2&   0.11& 0.94&    2.3&    3.5&   0.11& 0.94&    2.3\\
NGC 6304 &    3.0&   1.58& 0.32&    0.9&    3.0&   1.49& 0.34&    0.9&    3.4&   1.39& 0.42&    1.2\\
NGC 6316 &    3.9&   0.72& 0.69&    1.6&    3.9&   0.66& 0.71&    1.6&    4.7&   0.51& 0.80&    1.6\\
NGC 6341 &   10.8&   0.43& 0.92&    9.9&   10.8&   0.53& 0.91&    9.9&   16.8&   0.53& 0.94&   14.5\\
NGC 6325 &    1.4&   1.04& 0.14&    1.1&    1.4&   1.02& 0.15&    1.2&    1.5&   1.02& 0.20&    1.3\\
NGC 6333 &    6.4&   0.88& 0.76&    4.4&    6.4&   0.89& 0.76&    4.4&    8.6&   0.89& 0.81&    7.1\\
NGC 6342 &    1.8&   0.63& 0.47&    1.5&    1.8&   0.58& 0.51&    1.5&    2.2&   0.35& 0.73&    1.9\\
NGC 6356 &    8.5&   2.96& 0.48&    4.7&    8.5&   2.94& 0.49&    4.8&   12.5&   2.94& 0.62&    7.1\\
NGC 6355 &    1.4&   0.64& 0.38&    1.4&    2.2&   0.09& 0.93&    1.6&    2.2&   0.05& 0.95&    2.0\\
NGC 6352 &    4.2&   3.19& 0.13&    0.7&    4.2&   3.00& 0.16&    0.7&    5.6&   2.91& 0.31&    1.3\\
IC 1257  &   20.1&   0.69& 0.93&    7.2&   20.0&   0.74& 0.93&    7.1&   36.4&   0.74& 0.96&   15.9\\
Terzan 2 &    1.0&   0.13& 0.76&    0.4&    1.0&   0.12& 0.78&    0.7&    1.0&   0.12& 0.78&    0.8\\
NGC 6366 &    5.9&   2.24& 0.45&    2.0&    5.9&   2.18& 0.46&    2.0&    8.0&   2.15& 0.57&    3.3\\
Terzan 4 &    0.9&   0.18& 0.68&    0.7&    1.0&   0.12& 0.79&    0.7&    1.1&   0.12& 0.81&    0.8\\
BH 229   &    2.7&   0.04& 0.97&    2.1&    2.8&   0.03& 0.98&    2.1&    2.8&   0.02& 0.99&    2.7\\
FSR 1758 &   12.0&   3.31& 0.57&    5.2&   12.0&   3.28& 0.57&    5.4&   21.3&   3.28& 0.73&    8.3\\
NGC 6362 &    5.4&   2.48& 0.37&    3.3&    5.3&   2.44& 0.37&    3.3&    7.2&   2.44& 0.49&    4.3\\
Liller 1 &    0.8&   0.12& 0.75&    0.2&    0.8&   0.12& 0.75&    0.5&    0.9&   0.12& 0.77&    0.6\\
NGC 6380 &    2.4&   0.10& 0.92&    1.7&    2.4&   0.10& 0.92&    1.7&    2.5&   0.10& 0.92&    1.7\\
Terzan 1 &    2.8&   0.67& 0.62&    0.1&    2.8&   0.67& 0.62&    0.1&    3.1&   0.65& 0.65&    0.3\\
Pismis 26 &    3.3&   1.76& 0.30&    1.6&    3.3&   1.71& 0.31&    1.6&    3.9&   1.68& 0.40&    2.3\\
NGC 6388 &    4.2&   1.00& 0.61&    1.6&    4.2&   0.95& 0.63&    1.6&    4.9&   0.85& 0.71&    1.9\\
NGC 6402 &    4.7&   0.27& 0.89&    2.8&    4.6&   0.27& 0.89&    2.6&    5.4&   0.27& 0.90&    3.0\\
NGC 6401 &    2.0&   0.04& 0.96&    1.5&    2.0&   0.05& 0.95&    1.5&    2.0&   0.05& 0.95&    1.7\\
NGC 6397 &    6.5&   2.57& 0.43&    3.3&    6.5&   2.53& 0.44&    3.3&    9.2&   2.51& 0.57&    4.6\\
Pal 6    &    2.9&   0.04& 0.97&    2.2&    2.9&   0.03& 0.98&    2.2&    2.9&   0.03& 0.98&    2.7\\
NGC 6426 &   16.7&   3.28& 0.67&    7.0&   16.7&   3.26& 0.67&    7.5&   27.6&   3.26& 0.79&   13.8\\
Djorg 1  &    8.6&   1.06& 0.78&    1.0&    8.6&   1.03& 0.78&    1.0&   12.1&   1.01& 0.85&    3.0\\
Terzan 5 &    1.9&   0.22& 0.80&    1.1&    2.0&   0.22& 0.80&    1.0&    2.1&   0.22& 0.81&    1.0\\
NGC 6440 &    1.5&   0.05& 0.93&    1.2&    1.5&   0.05& 0.94&    1.2&    1.6&   0.04& 0.95&    1.3\\
NGC 6441 &    4.7&   1.43& 0.53&    1.4&    4.7&   1.29& 0.57&    1.4&    6.0&   1.13& 0.68&    1.6\\
Terzan 6 &    1.3&   0.17& 0.77&    0.5&    1.4&   0.16& 0.79&    0.7&    1.4&   0.16& 0.79&    0.7\\
NGC 6453 &    3.0&   0.08& 0.95&    2.2&    2.6&   0.21& 0.85&    2.2&    2.6&   0.21& 0.85&    2.2\\
NGC 6496 &    4.6&   2.35& 0.32&    2.4&    4.6&   2.28& 0.33&    2.4&    5.9&   2.27& 0.44&    3.3\\
Terzan 9 &    2.7&   0.27& 0.82&    1.6&    2.7&   0.27& 0.82&    1.4&    2.7&   0.27& 0.82&    1.5\\
Djorg 2  &    0.8&   0.50& 0.21&    0.4&    0.8&   0.50& 0.23&    0.4&    0.9&   0.50& 0.30&    0.5\\
\hline
  \end{tabular}
 \end{scriptsize}
  \end{center}
  \end{table*}

\begin{table*}
 \begin{center}
 \centerline{\small {\bf Table 2.} (Contd.)}
 \bigskip
 \begin{scriptsize}
 \label{t:par2}
 \begin{tabular}{|l|r|r|r|r||r|r|r|r||r|r|r|r|}\hline
  Name & $apo^{st}$ &$peri^{st}$&$ecc^{st}$&$Z_{max}^{st}$ & $apo^{-5}$ &$peri^{-5}$&$ecc^{-5}$&$Z_{max}^{-5}$&$apo^{-12}$ &$peri^{-12}$&$ecc^{-12}$&$Z_{max}^{-12}$\\
  &kpc &kpc  &   &kpc &kpc &kpc   &   &kpc&kpc&kpc&   &kpc\\\hline
NGC 6517 &    3.7&   0.23& 0.88&    2.4&    3.7&   0.23& 0.88&    2.2&    4.0&   0.22& 0.89&    2.3\\
Terzan 10 &    5.3&   0.57& 0.81&    3.7&    5.3&   0.58& 0.80&    3.9&    6.7&   0.58& 0.84&    5.4\\
NGC 6522 &    1.4&   0.42& 0.54&    1.1&    1.4&   0.41& 0.55&    1.1&    1.6&   0.41& 0.59&    1.3\\
NGC 6535 &    4.8&   0.80& 0.72&    2.0&    4.8&   0.70& 0.75&    2.0&    5.8&   0.61& 0.81&    2.3\\
NGC 6528 &    1.1&   0.23& 0.66&    0.9&    1.4&   0.10& 0.87&    0.9&    1.5&   0.10& 0.88&    0.9\\
NGC 6539 &    3.5&   1.85& 0.31&    2.5&    3.5&   1.90& 0.29&    2.6&    4.2&   1.90& 0.38&    3.0\\
NGC 6540 &    2.6&   1.14& 0.38&    0.5&    2.5&   1.11& 0.39&    0.5&    2.7&   1.05& 0.44&    0.8\\
NGC 6544 &    6.0&   0.43& 0.87&    3.1&    5.7&   0.48& 0.85&    3.1&    6.9&   0.48& 0.87&    4.6\\
NGC 6541 &    3.8&   1.29& 0.49&    2.4&    3.7&   1.29& 0.49&    2.4&    4.6&   1.29& 0.56&    3.0\\
ESO 280-06 &   13.7&   0.72& 0.90&   10.6&   13.7&   0.75& 0.90&   10.8&   22.1&   0.75& 0.93&   17.2\\
NGC 6553 &    3.9&   2.96& 0.13&    0.3&    3.9&   2.76& 0.17&    0.3&    4.9&   2.66& 0.30&    0.5\\
NGC 6558 &    1.7&   0.26& 0.72&    1.3&    1.7&   0.27& 0.72&    1.3&    1.7&   0.27& 0.73&    1.6\\
Pal 7    &    7.2&   3.78& 0.31&    0.7&    7.2&   3.62& 0.33&    0.8&   10.7&   3.59& 0.50&    1.7\\
Terzan 12 &    4.0&   1.87& 0.36&    1.2&    4.0&   1.73& 0.40&    1.1&    5.0&   1.55& 0.53&    1.4\\
NGC 6569 &    2.6&   1.46& 0.28&    1.3&    2.6&   1.44& 0.29&    1.3&    2.9&   1.41& 0.35&    1.6\\
ESO 456-78 &    3.2&   1.81& 0.27&    1.3&    3.2&   1.77& 0.28&    1.3&    3.8&   1.71& 0.37&    1.8\\
NGC 6584 &   18.6&   1.79& 0.82&   13.2&   18.6&   1.84& 0.82&   13.9&   34.2&   1.84& 0.90&   25.6\\
NGC 6624 &    1.7&   0.07& 0.92&    1.3&    1.8&   0.06& 0.94&    1.3&    1.9&   0.06& 0.94&    1.3\\
NGC 6626 &    3.2&   0.49& 0.73&    1.9&    3.2&   0.50& 0.73&    1.9&    3.5&   0.50& 0.75&    2.3\\
NGC 6638 &    2.7&   0.04& 0.97&    2.0&    2.5&   0.09& 0.93&    2.0&    2.5&   0.09& 0.93&    2.1\\
NGC 6637 &    2.4&   0.09& 0.93&    1.7&    2.4&   0.09& 0.93&    1.7&    2.4&   0.09& 0.93&    1.8\\
NGC 6642 &    2.2&   0.08& 0.93&    1.6&    2.3&   0.08& 0.93&    1.6&    2.3&   0.08& 0.93&    1.9\\
NGC 6652 &    3.6&   0.03& 0.98&    2.8&    3.5&   0.08& 0.96&    2.7&    3.5&   0.08& 0.96&    3.3\\
NGC 6656 &    9.8&   2.98& 0.53&    3.7&    9.8&   2.92& 0.54&    3.9&   16.1&   2.92& 0.69&    6.3\\
Pal 8    &    4.2&   0.87& 0.66&    1.7&    4.2&   0.82& 0.67&    1.7&    5.0&   0.69& 0.76&    1.7\\
NGC 6681 &    5.0&   0.48& 0.83&    4.8&    5.0&   0.49& 0.82&    4.8&    6.4&   0.49& 0.86&    5.7\\
NGC 6712 &    5.6&   0.05& 0.98&    4.4&    5.6&   0.04& 0.99&    4.4&    5.9&   0.04& 0.99&    5.7\\
NGC 6715 &   51.8&  14.41& 0.56&   46.8&   54.9&  14.48& 0.58&   46.9&  106.6&  14.48& 0.76&   78.7\\
NGC 6717 &    2.7&   0.64& 0.62&    1.4&    2.7&   0.66& 0.60&    1.4&    2.9&   0.66& 0.62&    1.8\\
NGC 6723 &    3.1&   1.68& 0.30&    3.1&    3.1&   1.68& 0.30&    3.0&    3.6&   1.68& 0.36&    3.5\\
NGC 6749 &    5.0&   1.47& 0.55&    0.3&    5.0&   1.40& 0.56&    0.3&    6.3&   1.32& 0.65&    0.7\\
NGC 6752 &    5.6&   3.46& 0.24&    2.1&    5.6&   3.33& 0.26&    2.1&    8.3&   3.31& 0.43&    3.5\\
NGC 6760 &    5.9&   1.94& 0.50&    0.6&    5.9&   1.84& 0.52&    0.7&    8.2&   1.77& 0.65&    1.6\\
NGC 6779 &   13.2&   0.71& 0.90&    9.3&   13.2&   0.74& 0.89&    9.4&   21.4&   0.74& 0.93&   14.1\\
Terzan 7 &   58.0&  14.34& 0.60&   53.2&   61.2&  14.43& 0.62&   53.4&  134.3&  14.43& 0.81&   77.8\\
Pal 10   &   11.0&   6.40& 0.26&    0.9&   11.0&   6.34& 0.27&    1.0&   19.5&   6.34& 0.51&    2.2\\
Arp 2    &   62.7&  17.69& 0.56&   60.3&   65.8&  17.78& 0.57&   53.9&  110.8&  17.78& 0.72&  102.9\\
NGC 6809 &    5.8&   1.18& 0.66&    4.7&    5.8&   1.20& 0.66&    4.7&    7.8&   1.20& 0.73&    6.3\\
Terzan 8 &   76.9&  17.48& 0.63&   69.0&   83.1&  17.57& 0.65&   66.2&  167.0&  17.57& 0.81&  149.2\\
Pal 11   &    8.7&   4.20& 0.35&    3.8&    8.7&   4.13& 0.35&    4.0&   13.9&   4.13& 0.54&    7.2\\
NGC 6838 &    7.3&   5.00& 0.18&    0.7&    7.3&   4.85& 0.20&    0.7&   11.2&   4.84& 0.40&    1.6\\
NGC 6864 &   16.0&   0.39& 0.95&   12.5&   16.0&   0.44& 0.95&   12.9&   25.6&   0.44& 0.97&   20.8\\
NGC 6934 &   42.7&   2.69& 0.88&   16.2&   44.3&   2.69& 0.89&   18.1&   66.1&   2.69& 0.92&   26.8\\
NGC 6981 &   22.0&   0.10& 0.99&   16.0&   21.8&   0.31& 0.97&   15.5&   36.1&   0.15& 0.99&   26.7\\
NGC 7006 &   53.3&   2.21& 0.92&   33.1&   55.4&   2.22& 0.92&   36.9&   81.9&   2.22& 0.95&   57.2\\
NGC 7078 &   10.8&   3.77& 0.48&    5.1&   10.8&   3.73& 0.49&    5.3&   17.7&   3.73& 0.65&    9.1\\
NGC 7089 &   19.0&   0.51& 0.95&   13.3&   18.9&   0.46& 0.95&   13.0&   29.1&   0.32& 0.98&   17.0\\
NGC 7099 &    8.5&   1.00& 0.79&    7.2&    8.5&   1.00& 0.79&    7.2&   12.4&   1.00& 0.85&   10.0\\
Pal 12   &   59.0&  15.01& 0.59&   50.9&   62.4&  15.12& 0.61&   51.2&  110.2&  15.12& 0.76&   76.2\\
Pal 13   &   71.2&   6.53& 0.83&   65.5&   74.8&   6.61& 0.84&   65.6&  153.4&   6.61& 0.92&   83.7\\
NGC 7492 &   26.1&   1.72& 0.88&   25.9&   26.4&   1.82& 0.87&   25.7&   41.0&   1.82& 0.91&   32.6\\
\hline
  \end{tabular}
 \end{scriptsize}
  \end{center}
  \end{table*}

As initial astrometric data for studying the orbital
motions of 152 GCs from the list of Vasiliev (2019)
we used the latest highly accurate observational data
from the Gaia satellite to date. These mostly include
the proper motions calculated using the EDR3 catalogue
by Vasiliev and Baumgardt (2021). In addition,
we used the new mean highly accurate heliocentric
distances from Baumgardt and Vasiliev (2021).

We numerically integrated the orbits for 5 and
12 Gyr backward. A table with the orbital parameters
in static and evolving potentials is provided. A comparison
of the orbital parameters of GCs derived in
the static and evolving potentials showed that the discrepancy
between the parameters is significant in the
time interval $[0,-12]$~Gyr, while in the time interval
$[0,-5]$~Gyr the orbital parameters of GCs undergo,
on average, minor changes compared to the orbital
parameters derived in the static potential, which fit
well into the limits of the statistical uncertainties
caused by the errors in the data, i.e., on time scales
from the present time to $-5$~Gyr the Galactic potential
may be deemed constant.

\bigskip{\bf ACKNOWLEDGMENTS}

We are sincerely grateful to the anonymous referees
for their very interesting and useful remarks that
allowed the paper to be improved.

\begin{table*}
 \begin{center}
 \caption[]
 {\small\baselineskip=1.0ex
 Infall times of GCs onto the Galaxy in the evolving potential with changing masses and changing sizes of the
components}
  \bigskip
  \begin{small}
 \label{t:infall}
 \begin{tabular}{|l|r|r|l|r|r|l|r|r|}\hline
 Name& GS &$T_{\rm infall},$ & Name & GS &$T_{\rm infall}$, & Name &GS & $T_{\rm infall}$,\\
          &    &Gyr  &       &    & Gyr   &           &   & Gyr \\\hline
NGC 288   & GE & $-$12.818 &NGC 5053  & H99&  $-$12.909   &FSR 1758 &Seq & $-$12.941 \\
NGC 362   & GE & $-$12.989 &NGC 5272  & H99&  $-$12.855   &Djorg 1  & GE & $-$12.993 \\
Whiting 1 & Sgr& $-$11.777 &NGC 5286  & GE & $-$12.872    &NGC 6584 & GE & $-$12.820 \\
Pal 1     & D  & $-$12.712 &NGC 5466  & GE & $-$12.329    &NGC 6656 &D   & $-$12.983 \\
E 1       & HE &$-$12.043  &NGC 5694  & HE &$-$12.036     &NGC 6715 & Sgr& $-$12.905 \\
Eridanus  &HE  & $-$11.310 &IC 4499   &Seq & $-$12.612    &Terzan 7 & Sgr& $-$12.739 \\
NGC 2419  & Sgr& $-$12.404 &NGC 5824  & H99&  $-$12.359   &Arp 2    & Sgr& $-$12.027 \\
Pyxis     & HE &$-$12.807  &Pal 5     & H99& $-$12.852    &Terzan 8 & Sgr& $-$12.878 \\
NGC 2808  & GE & $-$12.837 &NGC 5897  & GE & $-$12.932    &Pal 11   & D  & $-$12.931 \\
E 3       & D  & $-$12.932 &NGC 5904  & GE & $-$12.711    &NGC 6934 & GE & $-$12.289 \\
Pal 3     &HE  & $-$11.682 &ESO 224-8 &D   & $-$12.715    &NGC 6981 & GE & $-$12.850 \\
NGC 3201  &Seq & $-$12.515 &Pal 14    & GE& $-$11.584     &NGC 7006 & GE & $-$12.225 \\
Crater    & HE &$-$12.180  &NGC 6101  & Seq &$-$12.623    &Pal 12   & Sgr& $-$12.129 \\
NGC 4147  & GE &$-$12.950  &NGC 6205  & GE& $-$12.993     &Pal 13   & GE & $-$12.471 \\
Rup 106   & H99& $-$12.706 &Pal 15    & GE& $-$12.100     &NGC 7492 & GE & $-$12.901 \\
NGC 4590  & H99& $-$12.996 &NGC 6341  & GE& $-$12.873     &  --     & -- & --        \\
NGC 5024  & H99& $-$12.987 &IC 1257   & GE& $-$12.635     &  --     & -- & --          \\\hline
\end{tabular}
\end{small}
\end{center}
\end{table*}

\bigskip \bigskip\medskip{\bf REFERENCES}{\small

1. N. Aghanim, Y. Akrami, M. Ashdown, et al. (Planck Collab.), Astron. Astrophys. 641, A6 (2020).

2. C. Allen and A. Santillan, Rev. Mex. Astron. Astrofis. 22, 255 (1991).

3. B. M. Armstrong, K. Bekki, and A. D. Ludlow, Mon. Not. R. Astron. Soc. 500, 2937 (2021).

4. J. A. Arnold, A. J. Romanowsky, J. P. Brodie, L. Chomiuk, L. R. Spitler, J. Strader, A. J. Benson, and D. A. Forbes, Astrophys. J. Lett. 736, L26 (2011).

5. A. T. Bajkova and V. V. Bobylev, Astron. Lett. 42, 567 (2016).

6. A. T. Bajkova and V. V. Bobylev, Research Astron. Astrophys. 21, 173 (2021a). 

7. A. T. Bajkova and V. V. Bobylev, Astron. Rep. 65, 737 (2021b). 

8. A. T. Bajkova, G. Carraro, V. I. Korchagin, N. O. Budanova, and V. V. Bobylev, Astrophys. J. 895, 69 (2020).

9. G. Battaglia, S. Taibi, G. F. Thomas, and T. K. Fritz, astro-ph/2106.08819 (2021).

10. H. Baumgardt and E. Vasiliev, astro-ph/2105.09526 (2021).

11. K. Bekki, M. A. Beasley, J. P. Brodie, and D. A. Forbes, Mon. Not. R. Astron. Soc. 363, 1211 (2005).

12. M. Bellazzini, Mon. Not. R. Astron. Soc. 347, 119 (2004).

13. P. Bhattacharjee, S. Chaudhury, and S. Kundu, Astrophys. J. 785, 63 (2014).

14. J. Bland-Hawthorn and O. Gerhard, Ann. Rev. Astron. Astrophys. 54, 529 (2016).

15. V. V. Bobylev and A. T. Bajkova, Astron. Lett. 42, 1 (2016).

16. J. S. Bullock and K. V. Johnston, Astrophys. J. 635, 931 (2005).

17. C. A. Correa, J. S. B. Wyithe, J. Schaye, and A. R. Duffy, Mon. Not. R. Astron. Soc. 452, 1217 (2015)

18. T. Garrow, J. J. Webb, and J. Bovy, Mon. Not. R. Astron. Soc. 499, 804 (2020).

19. F. A. G\'omez, A. Helmi, A. G. A. Brown, and Y.-S. Li, Mon. Not. R. Astron. Soc. 408, 935 (2010).

20. H. Haghi, A. H. Zonoozi, and S. Taghavi, Mon. Not. R. Astron. Soc. 450, 2812 (2015).

21. W. Harris, astro-ph/1012.3224 (2010).

22. A. Helmi, F. van Leeuwen, P. J.McMillan, et al. (Gaia Collab.), Astron. Astrophys. 616, 12 (2018).

23. A. Irrgang, B. Wilcox, E. Tucker, and L. Schiefelbein, Astron. Astrophys. 549, 137 (2013).

24. I. King, Astrophys. J. 67, 471 (1962).

25. H. H. Koppelman and A. Helmi, astroph/2006.16283 (2020).

26. D. Massari, H. H. Koppelman, and A. Helmi, Astron. Astrophys. 630, L4 (2019).

27. M. Miyamoto and R. Nagai, Publ. Astron. Soc. Jpn. 27, 533 (1975).

28. J. F. Navarro, C. S. Frenk, and S. D. M. White, Astrophys. J. 490, 493 (1997).

29. A. Perez-Villegas, B. Barbuy, L. O. Kerber, S. Ortolani, S. O. Souza, and E. Bica, Mon. Not. R. Astron. Soc. 491, 3251 (2020).

30. J. L. Sanders, E. J. Lilley, E. Vasiliev, N. W. Evans, and D. Erkal, Mon. Not. R. Astron. Soc. 499, 4793 (2020).

31. R. Sch\"onrich, J. Binney, and W. Dehnen, Mon. Not. R. Astron. Soc. 403, 1829 (2010).

32. S. Trujillo-Gomez, J. M. D. Kruijssen, M. Reina-Campos, J. L. Pfeffer, B. W. Keller, R. A. Crain, N. Bastian, and M. E. Hughes, Mon. Not. R. Astron. Soc. 503, 31 (2021).

33. E. Vasiliev, Mon. Not. R. Astron. Soc. 484, 2832 (2019).

34. E. Vasiliev and H. Baumgardt, astro-ph/2102.09568 (2021).

35. W. Wang, J. Han, M. Cautun, Z. Li, and M. Ishigaki, South Calif. Publ. Manag. Assoc. 63, 109801 (2020).
}
  \end{document}